# Achieving Sub-Exponential Speedup in Gate-Based Quantum Computing for Quadratic Unconstrained Binary Optimization

*Tseng,Ying-Wei[1] ·Kao,Yu-Ting[1]·Chang,Yeong-Jar[1], Ou, Chia-Ho[2,3], Chang, Wen-Chih[2]*
*Industrial Technology Research Institute, Electro-Optical Systems Laboratory (EOSL)[1]*
*National Pingtung University[2]*
*Graduate School of Information Sciences, Tohoku University, Japan[3]*

## ABSTRACT

Recent quantum-inspired methods based on the Simulated Annealing (SA) algorithm have shown strong potential for solving combinatorial optimization problems. However, Grover's algorithm [1] in gate-based quantum computing offers only a quadratic speedup, which remains impractical for large problem sizes. This paper proposes a hybrid approach that integrates SA with Grover's algorithm to achieve sub-exponential speedup, thereby improving its industrial applicability.

In enzyme fermentation, variables such as temperature, stirring, wait time, pH, tryptophan, rice flour and so on are encoded by 625 binary parameters, defining the space of possible enzyme formulations. We aim to find a binary configuration that maximizes the active ingredient, formulated as a 625-bit QUBO which is generated by historical experiments and AI techniques. Minimizing the QUBO cost corresponds to maximizing the active ingredient. This case study demonstrates that our hybrid method achieves sub-exponential speedup through gate-based quantum computing.

## 1. INTRODUCTION

Combinatorial optimization is a fundamental challenge across many domains, particularly when the solution space grows exponentially with the number of binary variables. Quadratic Unconstrained Binary Optimization (QUBO) provides a flexible framework for modeling such problems, but solving large QUBO instances remains computationally intractable using conventional methods.

In 1996, Grover's quantum search algorithm demonstrated a theoretical quadratic speedup, reducing the search complexity from $O(N)$ to $O(\sqrt{N})$, where $N$ denotes the total number of candidates. However, this speedup is still insufficient for many real-world applications, particularly in cases where the number of variables in a QUBO problem is $n$ (typically ranging from 100 to 10,000). In such cases, $N = 2^n$ becomes so large that even $\sqrt{N}$ remains computationally infeasible.

D-Wave [2] was the first to efficiently solve QUBO problems using quantum hardware, though its solutions often exhibited higher error rates. Fujitsu [3] later introduced a quantum-inspired approach by leveraging SA with semiconductor technology to achieve lower error. Quantum-inspired method mimics quantum computing through repeated QUBO cost computation instances, and it served as a key inspiration for this work. While SA is widely adopted for combinatorial optimization, it becomes time-consuming in high-dimensional spaces. In contrast, Grover's algorithm in gate-based quantum computing provides a quadratic speedup, but the optimization time becomes impractical for larger number of variables (e.g., $n>50$). These limitations motivate the need for more realistic methods.

In this paper, we propose a hybrid optimization framework that integrates SA with Grover's algorithm to achieve sub-exponential acceleration. The approach is based on two key ideas: (1) comparing performance against SA rather than brute-force, accepting some loss in optimality for efficiency, and (2) adopting a $2^q$ parallel configuration instead of searching the entire $2^n$ space. For instance, when the parameter q=10 is chosen to be much smaller than n=625, and a gate-based quantum computing framework is employed to achieve $2^q$-level parallelism instead of $2^n$-level parallelism, the optimization process can be executed within a practically feasible timeframe. This configuration yields a speedup of $O(2^{q/2})$, indicating a sub-exponential enhancement compared to conventional SA methods.

The proposed framework is designed with three conceptual mechanisms—segmentation, quantum acceleration, and circuit simplification—which together bridge classical heuristic search and gate-based quantum computation.

We validate our method using a real-world case, where historical enzyme fermentation data and AI technologies are used to construct a 625-bit QUBO model. Experiments conducted in a Python environment demonstrate the sub-exponential speedup of our hybrid approach. The QUBO model used in this study was derived from our earlier works on enzyme fermentation optimization, conducted in collaboration with a biopharmaceutical partner to improve enzyme formulations. The $2^{625}$ search space in QUBO is vast, and each experiment takes over an hour, making manual optimization infeasible. Our first study [4] demonstrated that active ingredient levels could be enhanced with fewer experiments, while the second study [5] showed that the runtime of SA does not grow exponentially with the number of variables $n$. These findings highlight the industrial relevance of achieving further acceleration through quantum and quantum-inspired methods.

In parallel with these efforts, our team (Software Division, Electronic and Optoelectronic System, Industrial Technology Research Institute, Taiwan) has advanced a broader line of research encompassing fourteen quantum innovations since 2022, with three more scheduled for development in the coming year. While Foxconn is developing ion trap quantum hardware, we focus on software innovations that make quantum resources more accessible to end users, creating a complementary and synergistic ecosystem in Taiwan. These innovations span three domains:

(1) Quantum Circuit Simulation and Analysis.
- 2022 – Quantum polynomial representation enables: observable parallelism, independent derivation, clear analysis, and easier debugging.[6]
- 2024 – Quantum-Chiplet achieves exponential improvement in matrix-based design complexity [7, 8].
- 2026 – De-kron method aims to achieve exponential improvement in simulation time.

(2) Quantum Architecture and Circuit Design.
- 2022 – Quantum Monte Carlo circuits and analysis methods.[9]
- 2023 – Arbitrary logic design and RTL-to-Quantum automation.
- 2025/05 – Mixed-signal quantum circuit design (presented at SNUG Taiwan): combining analog simplification with digital flexibility and synthesizability, enabling massively parallel simulation of stock price fluctuations [10].
- 2025/08 – Random injection and payoff computation for financial risk asset analysis: demonstrated quantum advantage in real-world scenarios [11].

- 2025/11 – Application of VLSI design methodologies to quantum circuit design and analysis for Monte Carlo simulations and non-integer-point QFT.
- 2026/Q2 – Fully digital design flow for option pricing, enabling nonlinear computations previously considered intractable.

(3) Quantum Industrial Applications (Machine Learning & QUBO).
- 2024/H1 – Hybrid-QNN architecture design, solving the exponential input growth problem in quantum systems [12]
- 2024/H2 – AI + quantum-inspired optimization [4]
- 2025/06 – Quantum Accelerations [5]
- 2025/09 – Hybrid of SA and Grover (this work)
- 2026/Q4 – QUBO-to-RTL automation enabling both FPGA and gate-based quantum systems to rapidly solve combinatorial optimization problems.

This work contributes a new quantum-classical hybrid framework for large-scale QUBO optimization, demonstrated through a real-world industrial application. The case study highlights both the substantial advantage of our method when benchmarked against SA, and the sub-exponential speedup achieved by leveraging a $2^q$ parallel configuration rather than exploring the full $2^n$ search space. Empirical results show tangible performance gains when using gate-based quantum computing.

## 2. METHODOLOGY

### 2.1. Overview of the Hybrid SA–Grover Framework

The proposed framework combines Simulated Annealing (SA) with Grover's quantum search to accelerate the most computationally demanding stage of SA: the evaluation of the QUBO cost function across multiple candidate solutions.

To illustrate, we refer to the scenario depicted in **Figure 1**, in which classical SA requires 4096 sequential evaluations of the QUBO cost function. If parallelized across 1024 GPUs, this reduces to 4 outer iterations, each evaluating 1024 configurations simultaneously—referred to as the "$2^q$ configurations" approach with q = 10. Grover's algorithm offers a quantum analogue to this parallelism: instead of evaluating $2^q$ configurations directly, it reduces the search complexity to $O(2^{q/2})$, which is approximately 32 iterations for q = 10. Since Big O notation suppresses constant factors, the actual number of Grover iterations is more accurately represented as $k \times 2^{q/2}$, where $k$ is a constant. As a result, each outer iteration requires about $32k$ Grover steps, leading to a total of $4 \times 32 \times k$ quantum iterations. This localized use of Grover's search acts as a quantum accelerator within SA, targeting only the costly evaluation step without requiring full-scale quantum resources. The remainder of the SA process—candidate generation, acceptance probability, and annealing schedule—remains entirely classical.

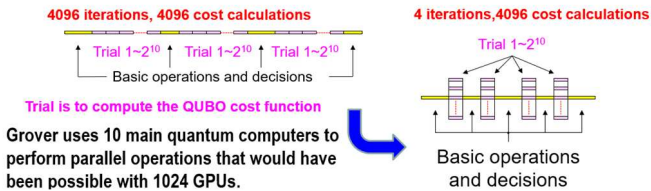

**Figure 1**. Iteration reduction using a 1024-configuration approach

To realize this hybrid computation, we design the SA–Grover hybrid framework composed of three core mechanisms: Segmentation, Acceleration, and Circuit Design Simplification as shown in Figure 2.

(1) Segmentation: The large n-bit problem ($n = 625$) is divided into smaller subproblems. During each SA iteration, most variables (n – q) are fixed, and only a small subset ($q = 2, 4, 6, \dots$) is explored by Grover's search. This selective search reduces the dimensionality handled by the quantum module.

(2) Acceleration: In conventional SA, each iteration must evaluate $2^q$ combinations, resulting in O($2^q$) cost. By embedding Grover's subroutine, the search cost is reduced to $O(2^{q/2})$, while the classical SA loop (temperature control and acceptance probability) remains intact. This hybridization converts theoretical quadratic speedup into a practical sub-exponential acceleration.

(3) Circuit Design Simplification: Because most variables ($n - q$) are fixed to 0 or 1, many quadratic terms in the QUBO expression vanish or reduce to constants. Consequently, the number of required qubits and quantum gates is significantly reduced, making the method feasible on current gate-based quantum hardware.

This three-part mechanism forms the basis of the proposed SA–Grover hybrid framework, balancing classical stability with quantum acceleration for industrial scenarios.

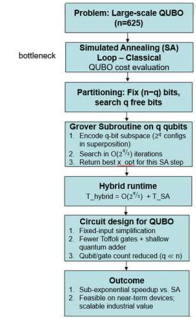

**Figure 2**. The proposed hybrid SA–Grover framework

### 2.2. Partitioning Strategy: Fewer Qubits for Large-Scale Problems

**Figure 3** illustrates the transition from classical Simulated Annealing (SA) to the proposed SA–Grover hybrid framework. On the left, the classical SA algorithm generates a candidate solution $X_{k+1}$ from the current state $X_k$, and evaluates its cost $f(X_{k+1})$. If the new cost is lower than $f(X_k)$, the candidate solution is accepted; otherwise, acceptance is probabilistically determined based on a random number and the thermal parameter of SA. This procedure requires one cost function evaluation per iteration.

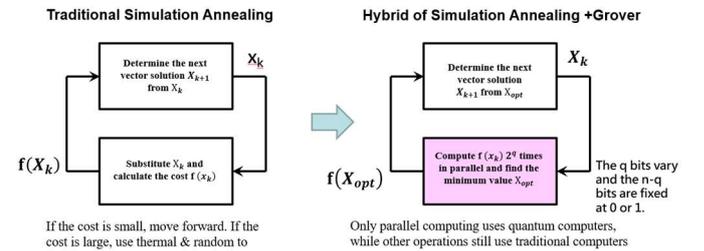

**Figure 3**. Transition from classical SA to the proposed hybrid SA–Grover framework.

The right side depicts the proposed SA–Grover framework. Since directly applying Grover's search over all *n*=625 variables is computationally infeasible, the search space is partitioned: only *q* variables are selected for quantum exploration, while the remaining *n–q* variables are fixed to either 0 or 1. Consequently, the quantum subroutine operates within a reduced *q*-dimensional subspace, preserving the global *n*-bit QUBO structure. Within each quantum subroutine, Grover's search evaluates all $2^q$ candidate configurations of the $q$ free bits in superposition and identifies the optimal candidate $X_{opt}$. This approach effectively replaces $2^q$ sequential cost evaluations in classical SA with $O(2^{q/2})$ Grover iterations, while the remainder of the SA process remains classical.

In this paper, we vary $q$=2, 4, 6, … 40, while keeping $n$=625, demonstrating that subspace-based quantum parallelization enables the use of limited qubit resources to tackle high-dimensional QUBO problems.

## 2.3 Quantum Circuit Design for the Simplified QUBO cost evaluation

In the previous section, we introduced a novel strategy for solving large $n$-bit QUBO problems by fixing $(n-q)$ variables and applying Grover's search only to the remaining $q$ free variables. While this approach may appear to simply reduce the search space, it also leads to a significant reduction in circuit complexity.

To illustrate, consider a 6-bit QUBO circuit. If three input bits are fixed—for example, $x_1 = 0$, $x_3 = 1$, and $x_5 = 0$—the circuit can be greatly simplified, as shown in Figure 4. In the first step, gate $G_1$, a NAND gate, has one input fixed at 0, so its output is forced to 1. Based on the similar method, $G_2$ and $G_3$ then depend only on $x_2$ and $x_4$, respectively. In the second step, G4, an OR gate, receives a fixed input of 1, which forces its output to 1 as well. This output becomes an input to G6, an AND gate, causing it to behave like a pass-through for its other input. Finally, the logic design is reduced to only one gate, G5, with only 2 Hadamard gates required.

This example demonstrates how fixing input variables can simplify logic gates and qubits significantly, improving the overall scalability and efficiency of quantum QUBO cost evaluations.

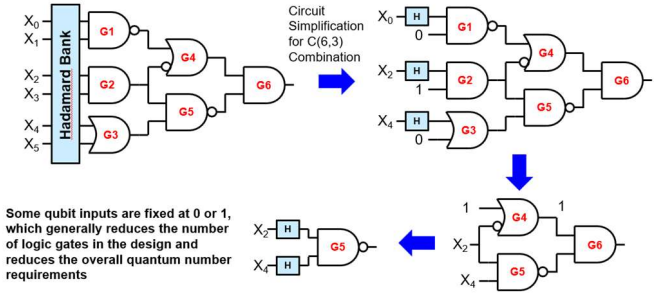

**Figure 4.** Circuit simplification by fixing input variables to 0 or 1

In addition to the aforementioned logic reduction, we examine this simplification phenomenon from the perspective of quantum circuit design. To illustrate, we consider a 5-bit QUBO example:

$$f(x_0, x_1, x_2, x_3, x_4) = x_1 x_2 + 2x_2 x_4 + 3x_1 x_4 + x_0 x_1 + 5x_0 + 2x_2 x_3 - 2x_2 + 4x_3 x_4 - 4x_4 + x_3 - 1$$

By fixing $x_0 = 0$ and $x_3 = 1$, the cost function reduces to three free variables $(x_1, x_2, x_4) = (x, y, x)$:

$$f(0, x, y, 1, z) = xy + 2yz + 3zx$$

This simplification reduces the search space from $2^5$=32 possible states to $2^3$=8, referred to as the "8 configurations". This enables 8 parallel computations of the QUBO cost, allowing the optimal solution to be identified using the Quantum-Grover algorithm. To leverage quantum speedup via Grover search, a quantum circuit must be designed to evaluate the QUBO cost. The reduced function can be implemented with three AND gates and a 2-bit adder as follows:
- Gate G1 computes the term $xy$;
- Gate G2 computes $yz$;
- Gate G3 computes $zx$.

The outputs of G1–G3 are then summed using a 2-bit adder to obtain the final cost value. The corresponding classical circuit implementation is illustrated in **Figure 5**.

In the quantum circuit, the AND operations are implemented using controlled-controlled-NOT (CCNOT) gates. As a result, G1, G2, and G3 correspond to three CCNOT gates. The fan-out mechanism is implemented by a CNOT gate. Finally, the 2-bit addition is performed using the quantum ripple-carry adder [13], completing the evaluation of the simplified 5-bit QUBO cost function, as shown in **Figure 6**.

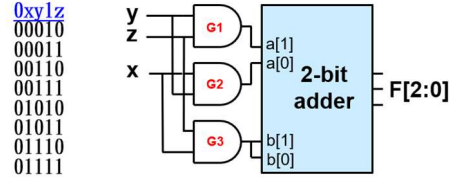

**Figure 5.** Classical circuit implementation of the simplified QUBO cost function using three AND gates (G1–G3) and a 2-bit adder

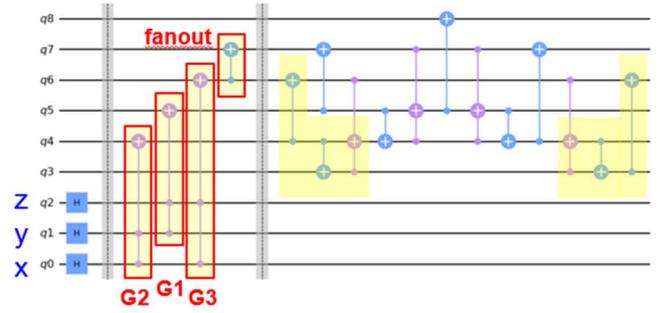

**Figure 6.** Quantum circuit mapping of the simplified QUBO cost function

Solving QUBO problems using quantum computing is often associated with quantum annealers such as D-Wave. However, this case demonstrates that gate-based quantum computing can also perform QUBO cost evaluations effectively. In particular, this highlights the capability of our method to solve larger QUBO problems with only a limited number of qubits.

## 2.4 Why Grover Can Be Improved from Quadratic to Sub-Exponential

**Table 1** compares several well-known computational paradigms, including Shor's algorithm, Monte Carlo (MC), Brassard's quantum amplitude estimation (QAE), brute-force search, Grover's algorithm, simulated annealing (SA), and our proposed hybrid SA–Grover framework. For all methods, the underlying search space is $N = 2^n$ for solving different problems.

**Table 1.** Comparison of quantum speedup methods including the proposed SA-Grover framework

| | Shor | MC | Brassard | Brutal | Grover | SA | Proposed |
|---|---|---|---|---|---|---|---|
| Search Space for n Binary Variables ($N$=$2^n$) | $2^n$ | $2^n$ | $2^n$ | $2^n$ | $2^n$ | $2^n$ | $2^n$ |
| # of parallelism | $2^n$ | 1 | $2^n$ | 1 | $2^n$ | 1 | $2^q$ |
| # of computations in classical computer | - | $O(m)$ | - | $O(N)$ | - | $O(2^q)$ | - |
| # of computations in quantum computer | $O(n)$ | - | $O(\sqrt{m})$ | - | $O(\sqrt{N})$ | - | $O(2^{q/2})$ |
| Advantages | Exp. | - | Quadratic | - | Quadratic | - | Sub-Exp. |

Shor's algorithm is well known for offering exponential speedup over the best classical algorithms, while the quadratic speedups offered by Brassard's and Grover's algorithms require a clear specification of their classical counterparts for comparison. For instance, Brassard's algorithm achieves a quadratic improvement over classical Monte Carlo methods, reducing the complexity from $O(m)$ to $O(\sqrt{m})$ for the same level of accuracy, where $m$ denotes the number of classical samples. Similarly, Grover's algorithm speeds up brute-force search by reducing the complexity from O($N$) to O($\sqrt{N}$), where $N$ is the size of the search space. Although both offer quadratic speedups with the same $2^n$ parallelism, their practical usability differs. Grover's algorithm becomes impractical for very large n (e.g., $n = 625$, so $N = 2^{625}$ and O($\sqrt{N}$) ≈ $10^{94}$). In contrast, Brassard's QAE builds a quantum estimator using q qubits such that $2^q \approx \sqrt{m}$ and $q \ll n$, thereby making the estimation process more scalable.

Inspired by this idea, the proposed hybrid SA-Grover algorithm uses a similar approach by designing a $2^q$ configuration such that $q \ll n$ to ensure the problem remains tractable. By redefining the classical baseline from brute-force search to SA, and by shifting the focus from $N$ to $q$, the speedup changes from O($\sqrt{N}$) to O($2^{q/2}$). This effectively transforms the theoretical quadratic speedup into a practical sub-exponential acceleration. Consequently, as the number of qubits increases, the computation time for the enzyme fermentation optimization task is reduced from 128 hours to 64, 32, 16, and eventually just 1 hour — demonstrating a significant quantum advantage that is both practically achievable and relevant to real-world applications.

## 3. EXPERIMENTAL RESULTS

### 3.1. Verification of the subroutine for simplified QUBO cost computation

To ensure the correctness of our implementation, the simplified QUBO cost computation subroutine is verified, as shown in Figure 7. The C(5,3) case refers to exploring $2^3=8$ input combinations out of the $2^5=32$ total candidates, by fixing $x_0=0$ and $x_3=1$ in the 5-bit QUBO function:

$$f(0, x, y, 1, z) = xy + 2yz + 3zx$$

The computed results from our subroutine match those obtained via manual calculation for all combinations of the explored bits, rather than all variables, confirming the functional correctness of both the QUBO cost subroutine and its integration into the hybrid SA–Grover framework.

**Figure 7.** Verification of the QUBO cost computation subroutine for the C(5,3) case

### 3.2 Runtime Analysis of the Baseline Method

We use PyQUBO as the baseline classical simulated annealing (SA) utility to evaluate speedup. In our experiments, the QUBO cost function was evaluated $10^5$ times using PyQUBO to measure both the total QUBO cost computation time and the overall runtime. The average per-call time was computed by dividing each of these totals by $10^5$, yielding the average QUBO cost computation time $t_Q$ and the average total runtime $t_{SA}$. The relationship is defined as:

$$t_{SA} = t_Q + t_{det}$$

where $t_{det}$ denotes the residual overhead unrelated to QUBO evaluation, measured to be approximately $1.3 \times 10^{-4}$ seconds.

For an $n$-bit QUBO problem, the time complexity of computing the QUBO cost scales as $t_Q = O(n^2)$, since it involves summing over all quadratic interactions in the binary vector. This complexity is empirically validated in Table 2, which reports the QUBO cost computation time $t_Q$ for various values of $n$. It is worth noting that, in our enzyme fermentation optimization case (625-bit QUBO), $t_Q = 0.228$ and $t_{SA} = 0.22813$ seconds.

Table 2. Empirical demonstration of O($n^2$) time complexity

| n | $t_Q$ (sec) | $t_Q/n^2$ |
|---|---|---|
| 5 | 3.05E-05 | 1.22E-06 |
| 10 | 7.78E-05 | 7.78E-07 |
| 15 | 1.51E-04 | 6.71E-07 |
| 20 | 2.54E-04 | 6.34E-07 |
| 25 | 3.90E-04 | 6.25E-07 |
| 625 | 2.28E-01 | 5.82E-07 |

### 3.3. Runtime Analysis of the Proposed Method

The proposed method is a hybrid SA-Grover framework. To assess the performance of the framework, the number of Grover qubits $q$ was varied from 2 to 40 in increments of 2. For each setting, we measured the QUBO evaluation time $T_Q$ and the total runtime $T_{SA}$.

To incorporate practical hardware considerations, we introduce a quantum overhead factor, denoted as $Q_{oh}$, which accounts for the additional latency introduced by Grover iterations, repeated quantum measurements, the speed gap between quantum and classical operations, and the overhead associated with classical-quantum data exchange.

The runtime of the baseline classical SA is given by:

$$T_{SA} = T_Q + T_{det}$$

In contrast, the runtime of the hybrid SA-Grover framework, which applies Grover's algorithm to a search subspace of size $2^q$, is expressed as:

$$T_{hy} = T_Q / (2^{q/2}) \times Q_{oh} + T_{det}$$

These equations indicate that quantum acceleration is applied exclusively to the QUBO cost evaluation step by Grover's algorithm, i.e., $T_Q$ is reduced. The remaining components of the SA process—namely, candidate generation, acceptance decisions, and the annealing schedule—are implemented entirely using classical computation, thus $T_{det}$ is unaffected.

Next, we performed a detailed runtime analysis by assigning specific parameter values to the hybrid SA–Grover framework. Since the total number of iterations required for SA to achieve a given optimization quality is approximately constant, we fixed this value as:

$$SA_{total} = 1 \times 10^{10}$$

This setup enables a direct comparison of runtime improvements across different values of $q$. The runtime of the QUBO cost evaluation was estimated as:

$$T_Q = SA_{total} \times t_Q = 2.28 \times 10^9$$

In comparison, the total runtime was slightly higher:

$$T_{SA} = SA_{total} \times t_{SA} = 2.2813 \times 10^9$$

where $t_Q$ and $t_{SA}$ are obtained from the previous section.

Table 3. Comparison of SA and hybrid SA-Grover framework

| q | 2 | 4 | 6 | 8 | 10 | 12 | 14 | 16 | 18 | 20 |
|---|---|---|---|---|---|---|---|---|---|---|
| QUBO time ($T_Q$) | 22800 | 22800 | 22800 | 22800 | 22800 | 22800 | 22800 | 22800 | 22800 | 22800 |
| Grover ($T_G$) | 1.14e+6 | 5.7e+5 | 2.85e+5 | 1.43e+5 | 71250 | 35625 | 17813 | 8906 | 4453 | 2227 |
| Speedup | 0.02 | 0.04 | 0.08 | 0.16 | 0.32 | 0.64 | 1.28 | 2.56 | 5.12 | 10.24 |
| SA time ($T_{sa}$) | 22813 | 22813 | 22813 | 22813 | 22813 | 22813 | 22813 | 22813 | 22813 | 22813 |
| Hybrid ($T_{hy}$) | 1.14e+6 +13 | 5.7e+5 +13 | 2.85e+5 +13 | 1.43e+5 +13 | 71263 | 35638 | 17826 | 8919 | 4466 | 2240 |

In Table 3, all runtime values are normalized over $10^5$ iterations. The table illustrates how the hybrid SA-Grover framework scales with increasing quantum resources specifically with qubits $q$=2, 4, 6, …20. This analysis confirms that the QUBO cost dominates the total runtime of SA, and it validates that accelerating this component using Grover's search is effective. However, to fully exploit Grover's quadratic speedup throughout the annealing process, the total number of SA iterations must remain significantly larger than the size of the quantum search subspace:

$$SA_{total} \gg 2^q$$

If this condition is not satisfied, the advantage of using multiple SA iterations to probabilistically escape local minima is lost. In such a case, the parallelism of quantum search becomes meaningless.

To ensure that the hybrid framework outperforms SA, the following condition must be satisfied:

$$2^{q/2} > Q_{oh}$$

In the present case, with $Q_{oh}$ =100, the equality becomes:

$$2^{14/2} = 2^7 = 128 > 100$$

indicating that the threshold for quantum advantage is reached at $q \geq 14$. Beyond this point, Grover's search outperforms purely classical cost evaluation. As $q$ increases further, the runtime decreases rapidly, demonstrating the expected sub-exponential acceleration. However, when $q$ becomes large, the runtime associated with the QUBO cost evaluation becomes negligible, and the remaining components of the SA process begin to dominate the total runtime. As a result, no further acceleration can be achieved. Therefore, the speedup saturates when q>34.

Figure 8(a) illustrates the runtime analysis. "QUBO" refers to the original QUBO computation time, while "Grover" indicates the reduced runtime achieved via Grover's algorithm. Similarly, "SA" denotes the total runtime of the classical SA process, and "SA+Grover" represents the improvement obtained by integrating Grover's search. Figure 8(b) presents the corresponding speedup results: "X_QUBO" represents the speedup factor, calculated as the runtime of "QUBO" divided by the runtime of "Grover". Similarly, "X_SA" denotes the speedup obtained by dividing the runtime of classical SA by that of the hybrid SA-Grover approach. Finally, for the 625-bit enzyme fermentation QUBO problem, the optimal quantum variable range is found to be $q$ = 14 to 34, where the trade-off between Grover acceleration and the classical residual yields the most effective sub-exponential speedup.

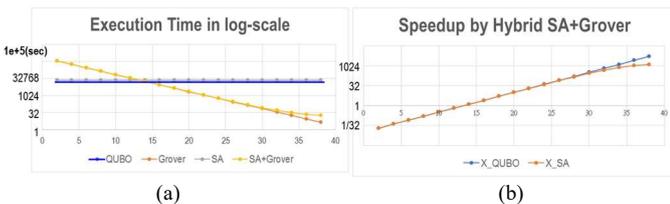

Figure 8. Comparison between Hybrid SA-Grover and Classical SA: (a) Runtime, (b) Speedup.

## 4. CONCLUSION

This work presents a hybrid SA–Grover framework that enables gate-based quantum computing to solve large-scale QUBO problems. While Grover's algorithm is often deemed impractical due to its theoretical quadratic speedup, we demonstrate that by redefining the performance baseline from brute-force to simulated annealing (SA), and by restricting the quantum search to a relevant subspace, this speedup can be transformed into a practical sub-exponential acceleration.

Assuming a conservative 100× quantum overhead—including slower quantum gate speeds, multiple Grover iterations, and quantum–classical data exchange—we observe clear runtime improvements: adding two qubits approximately halves the execution time. For a 625-bit enzyme fermentation problem, our method achieves speedups of 477×, 750×, and 1051× over traditional SA with 32, 34, and 36 qubits, respectively. We further show that fixing subsets of variables can significantly reduce the number of required qubits and quantum gates, thereby improving compatibility with near-term quantum hardware.

Overall, this study reframes the role of Grover's algorithm: not as a brute-force replacement, but as a scalable quantum enhancement to classical heuristics. This perspective paves the way for scalable quantum-enhanced combinatorial optimization and opens new opportunities for achieving quantum advantage in real-world applications such as biomanufacturing, logistics, and finance.